\documentclass[lettersize, journal]{IEEEtran}

\usepackage{cite}
\usepackage[affil-it]{authblk}
\usepackage[T1]{fontenc}
\usepackage[utf8]{inputenc}
\usepackage{graphicx}
\usepackage{amsmath, amsfonts, amssymb}
\usepackage{hyperref}
\usepackage{mathtools}
\usepackage{tabularx}
\usepackage{cancel}
\usepackage[RPvoltages]{circuitikz}
\usepackage{url}

\begin{document}
    \title{Synchronization of Josephson junctions in series array}
    \author{Abhijit~Bhattacharyya
    %\affil{Nuclear Physics Division \\ Bhabha Atomic Research Centre \\ Mumbai 400 094 \\ India}
    \thanks{Nuclear Physics Division, Bhabha Atomic Research Centre, Mumbai 400 085, India}
    \thanks{vega@barc.gov.in; abhihere@gmail.com}}

\maketitle

%%%%%%%%%%%%%%%%%%%%%%%%%%%%%%%%%%%%%%%%%
\begin{abstract}
Multi-qubit quantum processors coupled to networking provides the state-of-the-art quantum computing platform. However, each qubit has unique eigenfrequency even though fabricated in the same process. To continue quantum gate operations besides the detection and correction of errors it is required that the qubits must be synchronized in the same frequency. This study uses Kuramoto model which is a link between statistical mean-field technique and non-linear dynamics to synchronize the qubits applying small noise in the system. This noise could be any externally applied noise function or just noise from the difference of frequencies of qubits. The Kuramoto model tunes the coupled oscillators adjusting the coupling strength between the oscillators to evolve from the state of incoherence to the synchronized state.
\end{abstract}


%%%%%%%%%%%%%%%%%%%%%%%%%%%%%%%%%%%%%%%%%%%%%%

%%%%%%%%%%%%%%%%%%%%%%%%%%%%%%%%%%%%%%%%%%%%%%
\section{Introduction} \label{sec:intro}
\IEEEPARstart{J}osephson junction controls the flow of magnetic flux quanta through frequency and voltage.
% Controlling the flow of magnetic flux quanta, Josephson junction can maintain a nice relationship between frequency and voltage.
Modern instruments require measurement of voltage with a reproducible capability exceeding the uncertainty of realization of the SI volt (currently $0.4$ parts on $10^6$). Before 1972, $SI$ volt was represented by using carefully stabilised Weston cell banks~\cite{benz2004}. Drift and transportability problems with these electrochemical artifact standards limited the uniformity of voltage standards to about $1$ part in $10^6$. These uniformity was drastically improved by the usage of Josephson junction~\cite{benz2004}.

Josephson equation for supercurrent through a superconducting tunnel junction, called as {\emph{DC Josephson Effect}}, is defined as~\cite{josephson1961,josephson1964,josephson1973}
\begin{equation}
I = I_c \sin \left[\frac{4 \pi e}{h} \int V dt\right],  \label{eq:jDCeq}
\end{equation}
where $I_c$ is critical current, $h$ is Planck's constant and $e$ is electron charge. When a dc voltage is applied in equation \eqref{eq:jDCeq}, the phase will vary linearly with time and current will be sinusoidal with amplitude $I_c$ and frequency $f_J$ = $2eV/h$. The magnetic flux threading a superconducting loop or hole is quantized~\cite{deaver1961}. The superconducting magnetic flux quantum $\Phi_0$ = $h/(2e)$ is $2.0678 \times 10^{-15}$ $Wb$. The inverse of flux quantum $1/\Phi_0$ is called Josephson constant $K_J$ defined as $2e/h$ has a value of $483.597$ $GHz/mV$. During each oscillation, a single quantum of magnetic flux $h/(2e)$ passes through the junction which is very difficult to measure. However, if an alternating current with frequency $f$ is applied across the junction, there is a range of bias current for which flow of flux quanta will phaselock to the applied frequency. Under this phase locked condition, the average voltage across the junction is precisely $(h/2e)f$. This effect is known as {\emph{ac Josephson effect}} observed as a constant voltage step at $V$=$(h/2e)f$ in the $I-V$ characteristic curve. This means a Josephson junction can act as a ``Voltage to frequency converter''. It is also possible for the junction to phaselock with the harmonics of $f_J$ resulting in a series of steps at voltages $V$=$nf(h/2e)$, where $n$ is an integer denoting step number. This accuracy was limited to the condition that a Josephson voltage higher than $10 mV$ was never used~\cite{endo1983}. Therefore, if one obtain Josephson voltage over $100$ $mV$, the accuracy could be remarkably improved besides the ability to vary the Jsephson voltage with the frequency and step number could be utilized as potentiometer. Series array of Josephson junction~\cite{endo1983} has been effectively used in development of a potentiometer system to produce (1-10)V~\cite{benz2004, endo1983} with uncertainty about $2.5 \times 10^{-9}$~\cite{endo1983}. Larger series arrays were initially considered as impractical due to junction nonuniformity. The nonuniformity demanded each junction to be biased separately. In 1977, Levinsen {\emph{et al}}~\cite{levinsen1977} stated the important of the parameter $\beta_c$=$4 \pi e I_c R^2 C/h$ in determining the characteristics of RF induced Josephson steps. This $\beta_c$ is measure of the damping of Josephson oscillations by the junction shunting resistance $R$.

The Josephson junction is also a natural choice for submillimeter local oscillator~\cite{benz1991, wan1989} as one may capitalize the voltage controlled oscillator property. However, the disadvantage, in this case lies in very low power output. The Josephson constant clearly indicates that with dc voltage bias at $1$ $mV$ at $483.6$ $GHz$, the junction may accept $100$ $\mu A$ current keeping under the limitation of $I_c$ which limits the maximum output RF power at about $100$ $nW$. This requirement indicates series array of junctions with a common current bias demands  keeping all the junctions in phase.

However, the issue with series array of junctions operated with common current bias arises with nonuniformity of each junction due to fabrication processes~\cite{benz2004, endo1983}. When junctions are connected in series, the system behaves as a coupled oscillator and understanding the periodic solutions is important. Two special types of periodic solutions exist~\cite{swift1992}, namely, {\emph{in-phase state}} and {\emph{splay state}}.

An {\em in-phase} state with period $T$ is a state where all the oscillators always possess the same phase at all  times, {\em i.e.} $\theta_i(t) = \theta_j(t)$,  and $\theta_i(t+T) = \theta_i(t) + 2 \pi$.

The {\em splay-phase} or {\em anti-phase} or {\em rotating wave} state with period $T$ is a solution where the oscillators can be labeled so that $\theta_i(t)=\Theta(t+jT/N)$ for all $j$ for some function $\Theta(t+T)=\Theta(t)+2 \pi$. Thus, this state indicates that all the oscillators have the same waveform $\Theta(t)$ except for a shift in time. As per~\cite{swift1992}, one may imagine that each oscillator ``fires'' when it reaches a certain angle. For an {\em in-phase} solution, all the oscillators fire simultaneously at every instant $T$, while {\em splay-phase} state has a single oscillator firing every $T/N$ instant. Therefore, for {\em splay-phase} state, oscillators nearly coincide or coincide when $\dot{\theta}$ is small where as for large values, oscillators are not coherent. The definition of splay-phase does not imply that the phases of the oscillators are equi-spaced around he circle. The oscillators bunch up for smaller $\dot{\theta}$ while spread out for large $\dot{\theta}$. Therefore,  {\em splay-state} shows non-uniformity in the distribution of oscillators as they are coherent for smaller $\dot{\theta}$. It has been shown that~\cite{swift1992, swift1995}, the non-uniformity can be removed by determining a set of ``natural'' angles $\varphi_j$, so that the {\em splay-phase} solution satisfies $\varphi_j(t)=2 \pi j/N+2 \pi t/T+\text{const}$. The ``natural'' angle based dynamical system gets locked. This provides an idea of phase-locking $N$ oscillators, like $N$ Josephson junctions, having eigenfrequencies with smaller spread which may get locked to some resonating frequency.

Kuramoto model provides an exactly solvable mean-field model of coupled nonlinear oscillators connecting a large of them having distributed natural frequencies. This model links mean-field techniques and nonlinear dynamics together and also provides precise technique to tune the synchronization.

Section \ref{sec:KuramotoTheory} discusses the theory of the Kuramoto model, Section \ref{sec:KuramotoJJ} discusses on the reduction of the equations for the Josephson junctions connected in series to the Kuramoto Model framework and section \ref{sec:Analysis} discusses on the numerical analysis of the results for the generalised Kuramoto Model theory and Kuramoto model for Josephson junctions.
%%%%%%%%%%%%%%%%%%%%%%%%%%%%%%%%%%%%%%%%%%%%%%%%%%%%%%%%%%%%%%%%%%%%%%%%%%%%%%%%%

%%%%%%%%%%%%%%%%%%%%%%%%%%%%%%%%%%%%%%%%%%%%%%%%%%%%%%%%%%%%%%%%%%%%%%%%%%%%%%%%%
\section{Kuramoto Model}\label{sec:KuramotoTheory}

Let us consider a system of $N$ globally coupled differential equations with the stable limits cycles. Yoshiki Kuramoto developed a mathematical model for coupled oscillators ($n \geqslant 2$) to synchronize which is known as ``Kuramoto model''~\cite{kuramoto1986}.  In this model, each $j^{th}$ oscillator is represented by a phase variable $\theta_j(t)$, possessing its own natural frequency $\omega_j \in \mathcal{R}$. The dynamics of the system of coupled $N$ oscillators becomes
\begin{equation}
\dot{\theta_j}(t) = \omega_j +  \sum_{i = 1, j \ne i}^N K_{ji} \sin \left( \theta_j(t) - \theta_i(t)  \right), \ j \in \left\{ 1,\ldots, N \right \}, \label{eq:basicKura0}
\end{equation}
where $K_{ji}$ is coupling coefficient of the $j^{th}$ oscillator with all other oscillators in the system. Kuramoto assumed mean field coupling among phase oscillators such that $K_{ji} \approx K/N \geqslant 0$ where $K$ is mean coupling strength which changes \eqref{eq:basicKura0} as
\begin{equation}
\dot{\theta_j}(t) = \omega_j + \frac{K}{N} \sum_{i = 1, j \ne i}^N \sin \left( \theta_j(t) - \theta_i(t)  \right), \ j \in \left\{ 1,\ldots, N \right \}, \label{eq:basicKura1}
\end{equation}
where, $K \geqslant  0$ is the coupling strength among the oscillators whose frequencies are distributed with a probability density $g(\omega)$. One may find a suitable rotating frame like $\theta_j \rightarrow \theta_j - \Omega t$ transforming the system so that natural frequencies of the oscillators may have zero mean, where $\Omega$ is the first moment of the distribution function of natural frequencies $g(\omega)$. Therfore, one may consider the normal form calculation for the system such that one may define the system of equations as
\begin{equation}
\dot{\theta}_j=f_j(\theta_j)+\frac{K}{N} \sum_{i=1, i\ne j}^N g\left(\theta_i, \theta_j\right), \;\;  \theta_j \in \mathcal{R}^d, \,\, j=1,\ldots,N, \label{eq:coupledDynEq1}
\end{equation}
where, function $f_j(\theta_j)$ are eigenfrequencies defining the natural dynamics in the system. Here coupling parameter $K$ has been added with coupling strength $K/N$, $g$ is the phase response curve defining the interaction of the system. In the following section, we are not discussing with the stability of the dynamical system, bifurcation etc while one may consult other references like~\cite{guckenheimer2002}.

In the original paper~\cite{kuramoto1986}, Kuramoto considered the probability density $g(\omega)$ to be  uni-modal and symmetric centered at mean frequency $\overline{\omega}$  so that, without loss of generality, one can assume that the mean frequency $\overline{\omega}=0$ after a shift leading to $g(\omega) = g(-\omega)$ for the even and symmetric distribution $g(\omega)$.

To diagnose the feasibility of synchronization, Kuramoto introduced the order parameter $R(t)$ projecting the oscillation on unit circle where $R(t): 0 \leqslant R(t) \leqslant 1$ is a measure of the coherence of oscillators as
\begin{eqnarray}
    && R(t) e^{\jmath \psi(t)} = \frac{1}{N} \sum_{i=1}^N e^{\jmath \theta_i(t)},  \label{eq:orderParamKura}\\
    && \text{where } R(t)=\ 0 \text{ for asynchronised oscillators, } \nonumber \\
    && \text{ and } R(t) > 0 \text{ for synchronization}. \nonumber
\end{eqnarray}
The quantity $\psi(t)$ refers to average phase of all the oscillators at an instant $t$. Physically, this order parameter $R(t)$ is the centroid of a set of N points $e^{\jmath \theta_i}$ distributed in the unit circle in the complex plane at the instant $t$. If the phases are uniformly spread in the range $[-\pi, \pi]$, then $R \rightarrow 0$ indicates that the oscillators are not synchronized. All the oscillators become synchronized with the same average phase $\psi(t)$ for $R(t) \approx 1$. If the dynamics show stability of $R(t)$ at $1$, then the oscillators are synchronized and phaselocked. Eq. \eqref{eq:basicKura1} may be re-written by multiplying $Ke^{-\jmath \theta_j}$ on both sides of \eqref{eq:orderParamKura} and equating the imaginary parts of the both sides to reduce \eqref{eq:basicKura1} to
\begin{equation}
    \dot{\theta_j}(t) = \omega_j + K R(t) \sin \left( \psi(t) - \theta_j(t) \right) = v_j(\theta,\ \omega,\ t) \ \mbox{(say)}. \label{eq:basicKura2}
\end{equation}
Here, $v_j(\theta,\omega, t)$ is the angular velocity of a given oscillator with phase $\theta$ and natural frequency $\omega$ at the instant $t$. The equation \eqref{eq:basicKura2} reveals that the interaction is set through $R(t)$ and $\psi(t)$ while the phases $\theta_j$ seem to evolve independently from each other. Also the effective coupling is proportional to the order parameter $R(t)$ creating a feedback relation between coupling and synchronization. In the limit $K \rightarrow 0$,  \eqref{eq:basicKura2} reduces to
\begin{equation}
    \theta_j(t) \approx \omega_j t + \theta(0), \label{eq:oscSelf}
\end{equation}
where, $\theta_j(0)$ denotes initial phase of the $j^{th}$ oscillator and \eqref{eq:oscSelf} suggests that each oscillator oscillates with own natural frequencies  in the absence of coupling.

In the limit of infinite number of oscillators having a distribution of frequency, phase over time, Kuramoto described the system by the probability density $\rho \left( \theta, \omega, t \right)$ so that $\rho \left(\theta, \omega, t \right) d \theta$ gives the fraction of oscillators with phase between $\theta(t)$ and $\theta(t)+d \theta(t)$ at the instant $t$ for a given natural frequency $\omega$. Since $\rho$ is non-negative and $2 \pi$-periodic in $\theta$ satisfying the normalization condition
\begin{equation}
    \int_{-\pi}^{\pi} \rho \left( \theta, \omega, t \right) d \theta = 1. \label{eq:normaliz}
\end{equation}
The probability density function $g$ must also obey the equation of continuity using the angular velocity $v(\theta,\omega,t)$ as
\begin{eqnarray}
    & &\frac{\partial \rho(\theta, \omega, t)}{\partial t} + \frac{\partial }{\partial \theta} \left\{ \rho(\theta, \omega, t). v \right\} = 0,  \nonumber \\
    & & \frac{\partial \rho(\theta, \omega, t)}{\partial t} + \nonumber \\
    & & \frac{\partial}{\partial \theta} \left[
    \rho(\theta, \omega, t) \left\{\omega + K R(t) \sin \left(\psi(t) - \theta(t) \right) \right\}\right] = 0. \label{eq:continuity}
\end{eqnarray}
In the limit $R(t) \rightarrow 0$, the dynamics provides stationary solution for $\rho(\theta, \omega, t)=1/(2 \pi)$.

In the continuum limit,  \eqref{eq:orderParamKura} gets re-defined by the order parameter $R(t)$ and the average phase $\psi(t)$ incorporating previously described frequency distribution as
\begin{equation}
    R(t) e^{\jmath \psi(t)} = \int_{-\pi}^\pi \int_{-\infty}^{\infty} e^{\jmath \theta} \rho \left( \theta, \omega, t \right) g(\omega) d\omega d \theta. \label{eq:orderParam}
\end{equation}
In the strong coupling limit where $K \rightarrow \infty$ indicate $K \gg K_c$ where $K_c$ is critical coupling strength and \eqref{eq:basicKura2} reduces to system having phases reduced to the average phase as $\theta(t) = \omega t+\theta(0)=\psi(t)$.

From \eqref{eq:basicKura2}, if oscillators get into phaselocked condition, $v_i(t) \rightarrow 0$ which provides
\begin{equation}
    \omega_j = K R(t) \sin \left( \theta_j(t) - \psi(t) \right), -\frac{\pi}{2} \leqslant (\theta_j(t) - \psi(t)) \leqslant \frac{\pi}{2}. \label{eq:vZerolimit}
\end{equation}
From \eqref{eq:continuity}, partially synchronized state leading to a locked system can be described as $\frac{\partial}{\partial t} (\rho(\theta, \omega, t))=0$ which also means $\frac{\partial}{\partial \theta} \left(\rho(\theta, \omega, t).v(t)\right)=0$. Eq. \eqref{eq:vZerolimit}, in this partial synchronized state for $v_j(t) \rightarrow 0$ and $\frac{\partial}{\partial t} \left(\rho(\theta, \omega, t)\right)=0$, reduces to
\begin{equation}
\frac{\omega}{K R(t)} \rightarrow \sin(\theta_j(t) - \psi(t)), \nonumber
\end{equation}
which means
\begin{equation}
\rho(\theta, \omega, t) = \delta \left( \theta_j(t) - \psi(t) - \sin^{-1} \left( \frac{\omega}{KR(t)} \right) \right) H(\cos \theta), \label{eq:KuraCondition1}
\end{equation}
such that $\vert \omega \vert \leqslant KR(t)$ and
\begin{eqnarray}
H(x) =  &1,&\ x > 0, \nonumber \\
&0,& \mbox{elsewhere}..
\end{eqnarray}

Now, for the other condition $\frac{\partial}{\partial \theta} \left(\rho(\theta, \omega, t) v(t)\right)=0$ using  \eqref{eq:basicKura2},
\begin{eqnarray}
    & & \rho(\theta, \omega, t) v(t) = C \mbox{(say) = constant}, \nonumber \\
    \mbox{or, } & &  \rho(\theta, \omega, t) = \frac{C}{\vert \omega + KR(t) \sin(\theta_j(t) - \psi(t)) \vert}, \nonumber \\
    && \hskip 1cm  \vert \omega \vert \nleqslant  KR(t). \label{eq:KuraCondition21}
\end{eqnarray}
The constant $C$ can be determined from  \eqref{eq:normaliz} such that \eqref{eq:KuraCondition21} reduces to
\begin{eqnarray}
    && \rho(\theta, \omega, t) = \frac{\sqrt{\omega^2-K^2R^2(t)}}{2 \pi \vert \omega - KR(t) \sin(\theta_j(t) - \psi(t)) \vert}, \nonumber \\
    && \hskip 1cm  \vert \omega \vert \nleqslant  KR(t). \label{eq:KuraCondition22}
\end{eqnarray}

Therefore, the constraint on the probablity density of the oscillators may be
\begin{equation}
\rho(\theta, \omega, t)= \delta \left( \theta_j(t) - \psi(t) - \sin^{-1} \left( \frac{\omega}{KR(t)} \right) \right) H(\cos \theta), \nonumber
\end{equation}
\begin{equation}
\text{for } \vert \omega \vert \leqslant KR(t) \label{eq:KuraCond1}
\end{equation}
and
\begin{equation}
\rho(\theta, \omega, t) = \frac{\sqrt{\omega^2 - K^2R^2(t)}}{2 \pi \vert \omega - KR(t) \sin(\theta_j(t) - \psi(t)) \vert}, \text{elsewhere }. \label{eq:KuraCond2}
\end{equation}
Here $\delta$ is the Dirac delta function. Eqs. \eqref{eq:KuraCond1} and \eqref{eq:KuraCond2} indicate that partial synchronized states are divided into two groups depending on the natural frequencies. Oscillators having constraint $\vert \omega \vert \leqslant KR(t)$ operate in mean-field resulting in locking in a common average phase $\psi(t)=\Omega t$ where $\Omega$ is the average frequency of the ensemble of the oscillators in this regime. On the other side, the second group of oscillators having constraint $\vert \omega \vert > KR(t)$ rotate incoherently which are called as drifting oscillators.

Inserting \eqref{eq:KuraCond1} and \eqref{eq:KuraCond2} in \eqref{eq:orderParam} we get
\begin{eqnarray}
    R(t) &=& \int_{-\pi}^{\pi} \int_{-\infty}^{\infty} e^{\jmath(\phi(t)-\psi(t))} \nonumber \\
    & & \delta\left[\theta(t) - \psi(t) - \sin^{-1}\left(\frac{\omega}{KR(t)}\right) \right] g(\omega)d\theta d\omega \nonumber \\
    &+& \int_{-\pi}^{\pi} \int_{\vert \omega \vert \leqslant KR(t)} \frac{\sqrt{\omega^2 - K^2R^2(t)}g(\omega) d\theta d\omega}{2 \pi \vert \omega - KR(t) \sin(\theta(t) - \psi(t)) \vert}. \nonumber \\
    & & \label{eq:orderPKuraConds1}
\end{eqnarray}
Since $g(\omega)$ is even and symmetric, $g(\omega)=g(-\omega)$ and $\rho(\theta+\pi, -\omega)=\rho(\theta, \omega)$. The even function condition makes the second term of \eqref{eq:orderPKuraConds1} vanish which physically means all the incoherent oscillator solutions vanish resulting in order parameter $R(t)$ only for coherent synchronized oscillators that reform as
\begin{eqnarray}
    R(t) &=& \int_{\vert \omega \vert \leqslant KR(t)} \cos \left(\sin^{-1} \left(\frac{\omega}{KR(t)} \right) \right) g(\omega) d\omega d\theta, \nonumber \\
    &=& \int _{-\frac{\pi}{2}}^{\frac{\pi}{2}} \cos \theta g\left(KR(t)\sin\theta\right) KR(t) \cos \theta d \theta, \nonumber \\
    &=& KR(t) \int_{-\frac{\pi}{2}}^{\frac{\pi}{2}}\cos^2\theta g \left(KR(t) \sin \theta \right)d\theta. \label{eq:orderPKuraConds2}
\end{eqnarray}
Here, \eqref{eq:orderPKuraConds2} shows a trivial solution for which order parameter $R(t) = 0$ which actually shows incoherence as discussed earlier for $\rho\left( \theta, \omega, t\right)=1/(2 \pi)$. However, \eqref{eq:orderPKuraConds2} also suggests
\begin{equation}
    1 = K \int_{-\frac{\pi}{2}}^{\frac{\pi}{2}} \cos^2 \theta \; g\left( KR(t) \sin \theta \right) d \theta. \nonumber
\end{equation}
Setting $R(t)$ = $0$, considering $K$ = $K_c$ - the critical coupling strength we get,
\begin{equation}
 K_c = \frac{2}{\pi g(0)}, \label{eq:kc}
\end{equation}
that triggers the synchronization. In general, expanding the right hand side of \eqref{eq:orderPKuraConds2} in terms of powers of $KR(t)$ and considering $g^{\prime \prime}(0) <0$
the order parameter can be written as
\begin{equation}
    R(t) \sim \sqrt{-\frac{8\left( K - K_c \right)}{K_c^3 g^{\prime \prime}(0)}}, \label{eq:orderPKc}
\end{equation}
which shows that near the transition point, the order parameter \cite{ha2021, kuramoto1986}  yields the form $R(t) \sim (K - K_c)^{\beta}$ with $\beta\ =\ 1/2$ like second order phase transition.

The Kuramoto model can be generalized for a complex network including the connectivity parameter  in the coupling term as
\begin{equation}
    \dot{\theta}_j=\omega_j + \sum_{i=1}^N K_{ji} A_{ji} \sin(\theta_j-\theta_i), \label{eq:KuraGenNet}
\end{equation}
where, $K_{ji}$ is the coupling strength between nodes $j$ and $i$. $A_{ji}$ is the element of the adjacency matrix {$\mathbf{A}$} ($A_{ji} = 1$ if there is a connection between $j$ and $i$ else $A_{ji} = 0$ otherwise).

Any real system may have noise. Let us discuss on the effect of the noise for the Kuramoto model. The noise may arise from the variation of frequency of incoherent oscillators as they may not be identical or there may either be an external white noise or white noise inherent to the system. Therefore the model \eqref{eq:basicKura1} could be reframed as
\begin{eqnarray}
    \dot{\theta_j} &=& \sigma \omega_j + \frac{K}{N} \sum_{i=1}^N \sin\left( \theta_j (t) - \theta_i(t) \right) + \sqrt{\Gamma} \eta_j(t),  \nonumber \\
    &:& \;\; j \in \left\{1,\ldots,N\right\}, \label{eq:kuraNoise1}
\end{eqnarray}
where, both $\omega_j$ and $\eta_j(t)$ are Gaussian distributions having zero mean and unit variance while $\sigma$ and $\Gamma$ behave as amplitudes of the noise. Here last term refers to white noise in the system. Therefore \eqref{eq:kuraNoise1} physically indicates locally coupled oscillators having natural frequencies of oscillators  derived from Gaussian distribution in presence of stochastic effects like white noise due to fluctuations in the system. The reason for stochastic behavior may vary for different systems while any natural process exhibit stochastic behavior. .

The situation of $\lim \sigma\rightarrow 0$ refers to the Kuramoto model having identical oscillators in presence of gaussian white noise. The system behaves as if the system is in contact with a heat source and the dynamics is evolving in the statistical equilibrium.

The situation for  $\lim \Gamma \rightarrow 0$ indicates that the Kuramoto model has been constructed with oscillators having distributed natural frequencies in absence of gaussian white noise. The system behaves as nonlinear dynamical system relaxing to the non-equilibrium stationary state.

Beside this brief summary, one may also consult articles like \cite{dorfler2011}.

Next, let us transform the Josephson equations for series array of junctions to Kuramoto model.

%%%%%%%%%%%%%%%%%%%%%%%%%%%%%%%%%%%%%%%%%%%%%%%%%%%%%%%%%%%%%%%%%%%%%%%%%%%%%%%%%

%%%%%%%%%%%%%%%%%%%%%%%%%%%%%%%%%%%%%%%%%%%%%%%%%%%%%%%%%%%%%%%%%%%%%%%%%%%%%%%%%
\section{Kuramoto Model for Josephson junction series} \label{sec:KuramotoJJ}

The Josephson junction array can be constructed using Kirchhoff's laws considering each Josephson junction as a parallel circuit of two elements: an ideal resistance $\rho$  carrying ideal current $I_{\rho}$ and a junction carrying critical current $I_c$. Actual Josephson junction also contains a capacitor in parallel to the nonlinear inductor which we have neglected due to its very small value. Let each of $N$ junctions be connected serially and then coupled to external load having inductance $L$, resistance $R$ and capacitance $C$.

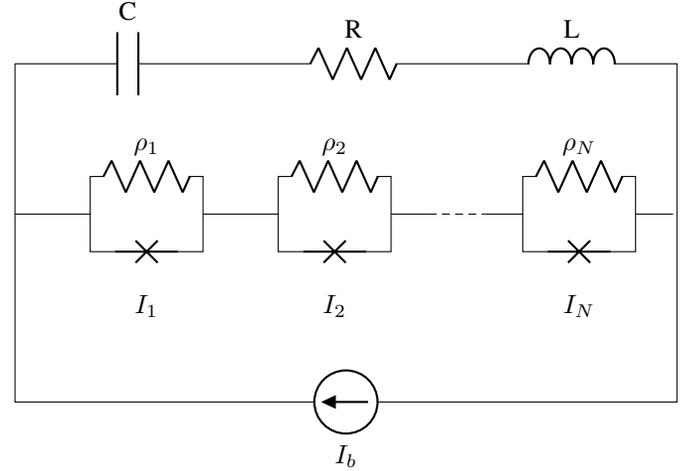
\begin{figure}[!h]
    \begin{center}
        \begin{circuitikz}[american]
            \draw
            (0,0)  to [C=C ]  ++(3, 0)
            to  [R=R] ++ (3, 0)
            to  [L=L]  ++(2.8, 0) %-- (7,-4)
            to  [short]  ++(0,-4.5)
            to [isource, l=$I_b$]  ++(-8.8, 0)
            to [short] (0, 0)
            ;

            \draw
            (0, -2) to [short] ++(0.5,0)
            ;

            \draw
            (0.5, -2) to [short] ++(0.5, 0)
            to [short]  ++(0, 0.5)
            to [R=$\rho_1$]  ++(1.5, 0)
            to [short]  ++(0, -1)
            to [barrier,l=$I_1$]  ++(-1.5,0)
            to [short]  ++(0, 0.5)
            ;
            \draw(2.5, -2) to [short] ++(0.5, 0)
            ;

            \draw
            (3, -2) to [short] ++(0.5, 0)
            to [short]  ++(0, 0.5)
            to [R = $\rho_2$]  ++(1.5, 0)
            to [short]  ++(0, -1)
            to [barrier, l = $I_2$]  ++(-1.5,0)
            to [short]  ++(0, 0.5)
            ;
            \draw(5.0, -2) to [short] ++(0.5, 0)
            ;

            \draw[densely dashed]  (5.5,-2) --  ++(0.75, 0)
            ;

            \draw
            (6.25, -2) to [short] ++(0.5, 0)
            to [short]  ++(0, 0.5)
            to [R = $\rho_N$]  ++(1.5, 0)
            to [short]  ++(0, -1)
            to [barrier, l = $I_N$]  ++(-1.5,0)
            to [short]  ++(0, 0.5)
            ;
            \draw(8.25, -2) to [short] ++(0.5, 0)
            ;
        \end{circuitikz}
        \caption[Qubit network]{Schematic circuit of qubits connected in series parallel to a Load.} \label{fig:cktParNetwork}
    \end{center}
\end{figure}

Let us consider Josephson junction in the series array, say $j^{th}$ junction and following Josephson equation; we  express the circuit shown in the Fig.\ref{fig:cktParNetwork} as
\begin{equation}
     \frac{V(t)}{\rho_j} + I_j \sin \phi_j + \frac{dQ}{dt} =I_b, \nonumber
\end{equation}
which can be written as,
\begin{equation}
     \frac{d \phi_j}{d t} = \frac{2 \pi \rho_j}{\Phi_0} \left(   I_b - I_j \sin \phi_j - \frac{dQ}{dt} \right). \label{eq:LCRcktV}
\end{equation}
Further,
\begin{eqnarray}
& & L \ddot{Q} + R\dot{Q} + \frac{Q}{C} = \sum_{k = 1}^N V_k, \nonumber \\
&\text{or, } & L \ddot{Q} + \left(R + \sum_{k=1}^N \rho_k \right)\frac{dQ}{dt} + \frac{Q}{C} = -\sum_{k = 1}^N I_k \rho_k \sin \phi_k,  \nonumber \\
& & \label{eq:LCRcktQ}
\end{eqnarray}
where $Q$ is the charge on load capacitor, $\Phi_0 = h/(2e)$ is magnetic flux quantum, $h$ is Planck's constant, $e$ being the charge of an electron. Here, junction resistance $\rho_k$ for any junction $k$ is very small compared to the load variable $Q/C$ such that one may consider, $Q/C-\sum_k \rho_k I_b$ $\approx$ $Q/C$. To understand the effect of external parameters like $L$, $C$ and $R$ on each junction, one may consider a scaled version of those parameters by choosing
\begin{equation}
    l = \frac{L}{N},\ r = \frac{R}{N}, \ c = NC. \label{eq:scaledLCR}
\end{equation}
Here, it is to be noted that
\begin{equation}
    \frac{\Phi_0}{I_j} = \frac{1}{2 \pi f_j^2 C_j} \nonumber
\end{equation}
where $f_j$ is the frequency and $C_j$ is the capacitance of $j^{th}$ junction.

Let us now consider transformation of time $t$ and charge $Q$ so that \eqref{eq:LCRcktV} and \eqref{eq:LCRcktQ} become dimensionless. From \eqref{eq:LCRcktV}
\begin{equation}
    \frac{\Phi_0}{2 \pi \rho_j I_j} \frac{d \phi_j}{d t} + \sin \phi_j + \frac{1}{I_j} \frac{dQ}{dt} =  \frac{I_b}{I_j} = \alpha_j. \nonumber
\end{equation}

Let us consider the following transformation relation to transform time $t$ to dimensionless form $\tau$ as
\begin{equation}
    \frac{\Phi_0}{2 \pi \rho_j I_j}\frac{d}{dt} \equiv \frac{d}{d \tau}. \label{eq:nondimTau}
\end{equation}
such that we may write
\begin{equation}
    \frac{d \phi_j}{d \tau} + \sin \phi_j + \frac{2 \pi \rho_j }{\Phi_0} \frac{dQ}{d \tau} = \alpha_j. \label{eq:tempPhiTau}
\end{equation}

Substituting dimensionless time $\tau$ and scaled parameters as in \eqref{eq:scaledLCR} in \eqref{eq:LCRcktQ} we get,
\begin{eqnarray}
    \frac{L}{N} \left(\frac{2 \pi \rho_j I_j}{\Phi_0}\right)^2 \frac{d^2 Q}{d \tau^2} &+& \frac{(R + \sum_{k=1}^N \rho_k)}{N} \left(\frac{2 \pi \rho_j I_j}{\Phi_0}\right) \frac{dQ}{d\tau} \nonumber \\
    &+& \frac{Q}{NC} = \frac{1}{N}\sum_{k = 1}^N - I_k \rho_k \sin \phi_k, \nonumber \\
    \text{or, } l\left(\frac{2 \pi \rho_j I_j}{\Phi_0}\right)^2 \frac{d^2 Q}{d \tau^2} &+& \left(r + \frac{\sum_{k=1}^N \rho_k}{N}\right) \left(\frac{2 \pi  \rho_j I_j}{\Phi_0}\right) \frac{dQ}{d\tau} \nonumber \\
    &+&  \frac{Q}{c} = -\frac{1}{N}\sum_{k = 1}^N  I_k \rho_k \sin \phi_k. \label{eq:L2l}
\end{eqnarray}
Let us also consider the following transformation to transform charge $Q$ to dimensionless form $q$ as
\begin{equation}
    \frac{2 \pi \rho_j I_j}{\Phi_0} Q \equiv q_j. \label{eq:nondimq}
\end{equation}
Therefore, through \eqref{eq:L2l}, \eqref{eq:LCRcktQ} transforms as
\begin{equation}
    \frac{d^2 q_j}{d \tau^2} + \gamma_j \frac{d q_j}{d \tau} + \omega_{0j}^2 q_j = - \frac{\delta_j}{N} \sum_{k=1}^N  I_k \rho_k \sin \phi_k. \label{eq:qjtau}
\end{equation}
Eq. \eqref{eq:nondimq} can be used to rewrite \eqref{eq:tempPhiTau} as
\begin{equation}
    \frac{d \phi_j}{d \tau} + \sin \phi_j + \epsilon_j \frac{dq_j}{d \tau} = \alpha_j, \label{eq:phijTau}
\end{equation}
where coefficients may be written as
\begin{eqnarray}
    \gamma_j & = & \left(\frac{\Phi_0}{2 \pi \rho_j I_j}\right)  \left(\frac{1}{l}\right) \left(r + \frac{\sum_{k=1}^N  \rho_k}{N}\right), \label{eq:gammaj} \\
    \omega_{0j}^2 & = & \left(\frac{\Phi_0}{2 \pi \rho_j I_j}\right)^2 \frac{1}{lc}, \label{eq:omega2oj} \\
    \delta_j & = & \left(\frac{\Phi_0}{2 \pi \rho_j I_j}\right)  \frac{1}{l}, \label{eq:betaj} \\
\text{and }    \epsilon_j & = & \frac{1}{I_j}. \label{eq:epsilonj}
\end{eqnarray}

Let us write the equation \eqref{eq:phijTau} in the uncoupled form for $\epsilon_j \rightarrow 0$ or $\dot{Q} \rightarrow 0$ such that we get,
\begin{equation}
    \frac{d \phi_j}{d \tau} = \alpha_j - \sin \phi_j. \label{eq:phijTauUncoupled}
\end{equation}

As discussed in the Section \ref{sec:intro}, the {\em splay-state} shows that transforming the dynamical system equations make a rigid system with coherent frequencies in weak coupling or uncoupled limit. Hence, let us transform $\phi_j$ in  \eqref{eq:LCRcktV} into `natural' angle $\psi_j$ such that $\frac{d \psi_j}{dt}=\text{constant}$. Eq. \eqref{eq:phijTauUncoupled} can be transformed in terms of the `natural' angle $\psi_j$   such that $d \psi_j/d t - c$, where $c$ is constant to be determined, {\em i.e.} transformation as  $\phi_j \rightarrow \psi_j$ as uniform rotation with first derivative remaining constant. The constant `$c$' may be determined with the fact that the time to complete one cycle by these two sets of coordinates must be same. Thus,
\begin{eqnarray}
    T &=& \int_0^T d\tau \nonumber  \\
      &=& \int_0^{2 \pi} \frac{d \psi_j}{c} = \int_0^{2 \pi} \frac{d \psi_j}{\omega_j} = \int_0^{2\pi} \frac{d \phi_j}{\left(\alpha_j - \sin \phi_j\right)}. \nonumber \\
    \text{or, } \frac{2 \pi}{\omega_j} &=& \frac{2 \pi}{\left(\sqrt{\alpha_j^2-1}\right)} , \text{  for $\alpha_j \geqslant 0$ i.e. $I_b\  \geqslant \  I_j$}, \nonumber
\end{eqnarray}

which shows
\begin{equation}
    \omega_j = \sqrt{\alpha_j^2-1}. \label{eq:omegajAlphaj}
\end{equation}

Then the transformation to the natural angles satisfies
\begin{equation}
    d\psi_j=\frac{\sqrt{\alpha_j^2-1}}{\alpha_j-\sin \phi_j}d\phi_j, \label{eq:dPsijdPhij}
\end{equation}
which on integration yields
\begin{equation}
    \psi_j = 2 \tan^{-1} \left(\sqrt{\frac{\alpha_j-1}{\alpha_j+1}} \tan \left(\frac{\phi_j}{2} + \frac{\pi}{4} \right) \right). \label{eq:naturalAnglexfrm}
\end{equation}

At this point, one may construct a transformation function $\psi(\phi_j)$ to translate any angle $\phi_j$ to its natural angle $\psi_j$ while another transformation function $\phi(\psi_j)$ may be used to invert as
\begin{eqnarray}
    & & \psi\left(\phi \right)= 2 \tan^{-1} \left(\sqrt{\frac{\alpha-1}{\alpha+1}} \tan \left(\frac{\phi}{2} + \frac{\pi}{4} \right) \right), \label{eq:func2normal}  \\
    & & \phi\left(\psi\right) = 2 \tan^{-1} \left(\sqrt{\frac{\alpha+1}{\alpha-1}} \tan \left(\frac{\psi}{2} \right) \right) - \frac{\pi}{2}. \label{eq:func2Invert}
\end{eqnarray}
Here, we use the shorthand: $\psi_j \equiv \psi(\phi_j)$ and $\phi_j \equiv \phi(\psi_j)$.

From \eqref{eq:naturalAnglexfrm},
\begin{equation}
\sin \phi_j = \frac{1- \alpha_j \cos \psi_j}{\alpha_j - \cos \psi_j}  = \alpha_j - \frac{\alpha_j^2-1 }{\alpha_j - \cos \psi_j}.  \label{eq:sinphipsi}
\end{equation}
Detailed derivation of \eqref{eq:sinphipsi} from \eqref{eq:naturalAnglexfrm} is shown in appendix \ref{sec:appendix1}.

Therefore, one may rewrite \eqref{eq:phijTau} using \eqref{eq:dPsijdPhij} and \eqref{eq:sinphipsi} as
\begin{eqnarray}
    \frac{d \psi_j}{d\tau} &=&  \frac{d \psi_j}{d \phi_j} \frac{d \phi_j}{d \tau} = \frac{\sqrt{\alpha_j^2 - 1}}{\alpha_j-\sin \phi_j}.\left(\alpha_j - \sin \phi_j - \epsilon_j \frac{dq_j}{d \tau} \right), \nonumber \\
    &=&  \sqrt{\alpha_j^2 - 1} - \frac{\epsilon_j \sqrt{\alpha_j^2-1}}{\alpha_j-\sin \phi_j}\frac{dq_j}{d\tau}. \label{eq:dpsijdtausqrt}
\end{eqnarray}
Let us rescale non-dimensional quantity $\tau$ as $\tilde{\tau}$ such that
\begin{eqnarray}
    & & \tau=\frac{\tilde\tau}{\sqrt{\alpha_j^2-1}} \nonumber \\
    &\implies& \frac{d}{d \tilde{\tau}} \equiv \frac{1}{\sqrt{\alpha_j^2-1}}\frac{d}{d \tau}   \nonumber \\
    &\implies&    \frac{d^2}{d \tau^2} \equiv \left(\alpha_j^2-1 \right) \frac{d^2}{d \tilde{\tau}^2}.  \label{eq:tau2tildeTau}
\end{eqnarray}
Eq. \eqref{eq:dpsijdtausqrt}, using \eqref{eq:tau2tildeTau}, transforms as
\begin{equation}
    \frac{d \psi_j}{d\tilde{\tau}} = 1 - \frac{\epsilon_j \sqrt{\alpha_j^2-1}}{\alpha_j-\sin \phi_j}\frac{dq_j}{d\tilde{\tau}}, \label{eq:weakCouplingdPsidtautilde}
\end{equation}
The weak-coupling solution of \eqref{eq:dpsijdtausqrt} may be written as
\begin{equation}
    \psi_j(\tau) \equiv \left(\sqrt{\alpha_j^2 -1}\right)\tau + c_j = \tilde{\tau} + \psi_{j0}, \label{eq:weakCouplingSolution1}
\end{equation}
where $c_j$ is the integration constant. Initially, at $\tau=0$, one may assume initial phase as $\psi_{j0}$ such that $c_j$=$\psi_{j0}$.
The reference \cite{swift1995} discusses about the importance of the weak coupling condition for the Josephson junction arrays and drift in $\psi_j$ may be obtained by averaging \eqref{eq:weakCouplingdPsidtautilde} over one cycle as
\begin{equation}
    \left \langle \frac{d \psi_j}{d\tilde{\tau}}\right \rangle = 1 - \frac{1}{2 \pi} \int_0^{2 \pi}\frac{\epsilon_j \sqrt{\alpha_j^2-1}}{\alpha_j-\sin \phi_j} \left(\frac{dq_j}{d\tilde{\tau}}\right) d\tilde{\tau}. \label{eq:weakCouplingdPsidtautildeAverage}
\end{equation}
Similarly, one may rewrite non-dimensional charge equation \eqref{eq:qjtau} in terms of $\tilde{\tau}$ as
\begin{eqnarray}
    \left(\alpha_j^2-1\right) \frac{d^2 q_j}{d \tilde{\tau}^2} &+& \gamma_j\sqrt{\alpha_j^2-1} \frac{d q_j}{d \tilde{\tau}} + \omega_{0j}^2 q_j \nonumber \\
    &=& - \frac{\delta_j}{N} \sum_{k=1}^N  I_k \rho_k \sin \phi_k. \label{eq:qjtautilde}
\end{eqnarray}
It is usually convenient to write $\sin(\phi_j)$=$\sin(\phi(\psi_j))$ in terms of its Fourier series as
\begin{eqnarray}
    \sin \phi(\psi_k) &=& \sum_{n=0}^\infty A_{kn} \cos \left(n \psi_{kn} \right) \nonumber \\
    &=& \sum_{n=0}^\infty A_{kn} \cos \left\{n \left(\tilde{\tau} + c_k\right)\right\}. \label{eq:sinPhiFourierForm}
\end{eqnarray}
Then \eqref{eq:qjtautilde} reduces to
\begin{eqnarray}
    \left(\alpha_j^2-1\right) \frac{d^2 q_j}{d \tilde{\tau}^2} &+& \gamma_j\sqrt{\alpha_j^2-1} \frac{d q_j}{d \tilde{\tau}} + \omega_{0j}^2 q_j \nonumber \\
    &=& - \frac{\delta_j}{N} \sum_{k=1}^N \sum_{n=0}^\infty  I_k \rho_k A_{kn} \cos\left\{n \left(\tilde{\tau} + c_k\right)\right\}.\nonumber \\ \label{eq:qjtautildeFourier}
\end{eqnarray}
One may obtain the steady-state solution of \eqref{eq:qjtautildeFourier} as
\begin{eqnarray}
    q_j(\tilde{\tau}) &=& -\frac{\delta_j}{N} I_k \rho_k B_{kn} \cos \left\{n \left(\tilde{\tau} + c_k\right) + \beta_{kn}\right\}, \label{qjsteady} \\
    \frac{dq_j(\tilde{\tau})}{d\tilde{\tau}} &=& \frac{\delta_j}{N} n  I_k \rho_k B_{kn} \sin \left\{n \left(\tilde{\tau} + c_k\right) + \beta_{kn}\right\}, \label{dqjdtautildesteady} \\
    \frac{d^2q_j(\tilde{\tau})}{d\tilde{\tau}^2} &=& \frac{\delta_j}{N} n^2  I_k \rho_k B_{kn} \cos \left\{n \left(\tilde{\tau} + c_k\right) + \beta_{kn}\right\}, \label{d2qjdtautilde2steady}
\end{eqnarray}
where
\begin{eqnarray}
    B_{kn}^2 &=& \frac{A_{kn}^2}{n^2 \gamma_j^2 \left(\alpha_j^2-1\right) + \left\{n^2 \left(\alpha_j^2-1\right) - \omega_{0j}^2\right\}^2}, \label{eq:Bkn} \\
    \beta_{kn} &=& \tan^{-1} \left[\frac{n \gamma_j \sqrt{\alpha_j^2-1}}{n^2 \left(\alpha_j^2-1\right)-\omega_{0j}^2}\right] = \beta_n. \label{eq:betakn}
\end{eqnarray}
Using the expression \eqref{eq:sinphipsi}, one may derive $A_{kn}$ and obtain
\begin{eqnarray}
    A_{k0} &=& \frac{1}{\pi} \int_{-\pi}^{\pi} \frac{1 - \alpha_k \cos \psi_k}{\alpha_k-\cos \psi_k} d \psi_k, \label{eq:Ak0}\\
    A_{kn} &=& \frac{1}{\pi} \int_{-\pi}^{\pi} \frac{1 - \alpha_k \cos \psi_k}{\alpha_k-\cos \psi_k} \cos  \left(\frac{n \pi \psi_k}{\pi}\right) d \psi_k \label{eq:Akn} \\
    & & \text{where }  n \neq 0. \nonumber
\end{eqnarray}
$B_{kn}$ denotes the amplitude of the linear damped oscillator while $\beta_{kn}$ denotes its phase. Therefore, $B_{kn}$ must be chosen to be positive.

Now, \eqref{eq:weakCouplingdPsidtautildeAverage} may be re-written as
\begin{eqnarray}
    \left \langle \frac{d \psi_j}{d\tilde{\tau}}\right \rangle &=& 1 - \frac{\epsilon_j \delta_j \sqrt{\alpha_j^2-1}}{2 \pi N} \int_0^{2 \pi} \left(\frac{1}{\alpha_j-\sin \phi_j} \right.\nonumber \\
    &\times& \left.\sum_{k=1}^N \sum_{n=0}^\infty n  I_k \rho_k B_{kn} \sin \left\{n \left(\tilde{\tau} + c_k\right) + \beta_{kn}\right\}\right) d\tilde{\tau}. \nonumber \\
    & & \label{eq:psitautildeFourier1}
\end{eqnarray}
Using \eqref{eq:sinphipsi},
\begin{eqnarray}
    & & \sin \phi_j =  \alpha_j - \frac{\alpha_j^2-1}{\alpha_j - \cos \psi_j}, \nonumber \\
    & & \text{or, }\alpha_j - \sin \phi_j = \frac{\alpha_j^2 - 1}{\alpha_j- \cos \psi_j}. \label{eq:phi2psij}
\end{eqnarray}
With this \eqref{eq:psitautildeFourier1} may be modified using \eqref{eq:phi2psij} to
\begin{equation}
    \text{or, } \left \langle \frac{d \psi_j}{d\tilde{\tau}}\right \rangle = 1 + \frac{K_j}{N} \sum_{k=1}^N A_k \sin \left(c_j - c_k - \zeta_j \right),     \label{eq:psiKuramotoGeneral0}
\end{equation}
where,
\begin{eqnarray}
    K_j &=& \frac{\epsilon_j \delta_j}{\sqrt{\alpha_j^2-1}\sqrt{\gamma_j^2 \left(\alpha_j^2-1\right)^2+\left(\omega_{0j}^2-\left(\alpha_j^2-1\right)^2\right)^2}}, \nonumber \\
    & & \label{eq:KGen} \\
    A_K &=& I_k \rho_k \left(1-\alpha_k^2+\alpha_k\sqrt{\alpha_k^2-1}\right), \label{eq:akGen} \\
    \zeta_j &=& \tan^{-1}\left(\frac{\gamma_j \sqrt{\alpha_j^2-1}}{\alpha_j^2-1-\omega_{0j}^2}\right) = \beta_{1j}. \label{eq:deltaGen}
\end{eqnarray}
Reader may check detailed description of the derivation in the appendix \ref{sec:appendix2}.

In the final step, one may replace the `initial values' of phases by their slowly evolving components like $\langle  \psi_j(\tilde{\tau}) \rangle$ and $\langle \psi_k(\tilde{\tau}) \rangle$. Also one may get firstorder averaged equation by dropping the angular brackets so that \eqref{eq:psiKuramotoGeneral0} transforms to
\begin{equation}
    \frac{d \psi_j}{d\tilde{\tau}} = 1 + \frac{K_j}{N} \sum_{k=1}^N A_k \sin \left(\psi_j(\tilde{\tau}) - \psi_k(\tilde{\tau}) - \delta \right). \label{eq:psiKuramotoGeneral}
\end{equation}

Eq. \eqref{eq:psiKuramotoGeneral} resembles the Kuramoto model in a generalized form. For the sake of mathematical formalities, it is important to note  that except terms corresponding to $n=1$ terms for other values of $n$ becomes zero.

To arrive at \eqref{eq:psiKuramotoGeneral}, it was assumed that the fabrication process may not guarantee exactly same values of parameters for each junction and hence one may consider that each junction has different internal resistance and different critical current. The difference may be very small for junctions prepared in the same batch. If the fabrication process is done in very skilled sequence \eqref{eq:psiKuramotoGeneral} may turn into special form for assuming $\rho_1=\rho_2=\ldots = \rho_N=\rho$ (say) and $I_1=I_2=\ldots=I_N=I_c$(say) so that each junction has nearly same frequency $f$ (say). This case of identical junctions has been studied extensively in may literatures.

The transformation \eqref{eq:nondimTau} for time leads to
\begin{equation}
    \frac{\Phi_0}{2 \pi \rho I_c}\frac{d}{dt} \equiv \frac{d}{d \tau}. \label{eq:nondimTauIdentical}
    %\frac{1}{4 \pi^2 f^2 C_J \rho}\frac{d}{dt} \equiv \frac{d}{d \tau}, \label{eq:nondimTauIdentical}
\end{equation}
while \eqref{eq:nondimq} entails
\begin{equation}
    \frac{2 \pi \rho I_c}{\Phi_0} Q \equiv q. \label{eq:nondimqIdentical}
    %\frac{4 \pi^2f^2\rho C_J}{I_c}Q \equiv q, \label{eq:nondimqIdentical}
\end{equation}

Consequently, \eqref{eq:qjtau} reduces to
\begin{equation}
    \frac{d^2q}{d\tau^2} + \gamma \frac{d q}{d\tau} + \omega_0^2 q = -\frac{\beta}{N} \sum_{k=1}^N \sin \phi_k, \label{eq:qjtauidentical}
\end{equation}
where
\begin{eqnarray}
    \gamma & = & \left(\frac{\Phi_0}{2 \pi \rho I_c}\right)  \left(\frac{1}{l\rho}\right) \left(r + \rho \right), \label{eq:gamma} \\
    \omega_0^2 & = & \left(\frac{\Phi_0}{2 \pi \rho I_c}\right)^2 \frac{1}{lc}, \label{eq:omega2o} \\
    \beta & = & \left(\frac{\Phi_0}{2 \pi \rho I_c}\right)  \frac{1}{l}, \label{eq:beta} \\
    %\gamma &=& \frac{(r+\rho)}{4 \pi^2 f^2 \rho C_J l} = \frac{\Phi_0}{2 \pi} \left(\frac{r+\rho}{\rho}\right)\left(\frac{1}{lI_c}\right), \label{eq:gammaIdentical} \\
    %\omega_0^2 &=& \frac{1}{\left(4 \pi^2 f^2 \rho C_J\right)^2 c} = \left(\frac{\Phi_0}{2 \pi}\right)^2 \left(\frac{1}{I_c^2\rho^2 c}\right). \label{eq:omega2Identical}
\end{eqnarray}
Eqs. \eqref{eq:Bkn} and \eqref{eq:betakn} become
\begin{eqnarray}
    & & B_n^2 = \frac{A_n^2}{n^2 \gamma^2 \left(\alpha^2-1\right) + \left\{n^2 \left(\alpha^2-1\right) - \omega_0^2\right\}^2}, \label{eq:BknIdentical} \\
    & & \beta_n = \tan^{-1} \left[\frac{n \gamma \sqrt{\alpha^2-1}}{\omega_0^2 - n^2 \left(\alpha^2-1\right)}\right]. \label{eq:betaknIdentical}
\end{eqnarray}
Repeating the earlier exercise, one may obtain the final phase equation \eqref{eq:psitautildeFourier1} as
\begin{eqnarray}
    \left\langle \frac{d\psi_j}{d\tilde{\tau}} \right\rangle &=& 1 - \frac{\beta}{2 \pi N \sqrt{\alpha^2-1}} \int_0^{2\pi} \left(\alpha-\cos \left( \tau + c_j\right)\right)\nonumber \\
    & & \times \sum_{k=1}^N\sum_{n=0}^\infty n B_n \sin \left\{n \left(\tilde{\tau} + c_k\right)+\beta_n\right\}d \tilde{\tau}. \nonumber \\
    & & \label{eq:psiFinalIdentical1}
\end{eqnarray}
In the case of identical junctions, computation shows that only $B_1$ exists while others are evaluated to zero. Thus,  \eqref{eq:psiFinalIdentical1} becomes
\begin{eqnarray}
    \left\langle \frac{d\psi_j}{d\tilde{\tau}} \right\rangle &=& 1 - \frac{B_1 \beta}{2 \pi N \sqrt{\alpha^2-1}} \int_0^{2\pi} \left(\alpha-\cos \left( \tau + c_j\right)\right) \nonumber \\
    &\times& \sum_{k=1}^N\sum_{n=0}^\infty n \sin \left\{n \left(\tilde{\tau} + c_k\right)+\beta_n\right\}d \tilde{\tau}. \nonumber
\end{eqnarray}
Integrating, we get
\begin{equation}
    \frac{d\psi_j}{d\tilde{\tau}}  = 1 + \frac{K}{N}
    \sum_{k=1}^N \sin \left(\psi_j(\tilde{\tau}) - \psi_k(\tilde{\tau}) - \beta_1 \right), \label{eq:psiFinalIdentical2}
\end{equation}
where
\begin{equation}
    K = \frac{\pi B_1 \beta}{2\pi \sqrt{\alpha^2-1}}. \label{eq:KValue}
\end{equation}
Eq. \eqref{eq:psiFinalIdentical2} exactly resembles as the Kuramoto model.

In the following section, let us try to understand general characteristics of the Kuramoto model in general and in the context of Josephson junction array.

%%%%%%%%%%%%%%%%%%%%%%%%%%%%%%%%%%%%%%%%%%%%%%%%%%%%%%%%%%%%%%%%%%%%%%%%%%%%%%%%%

%%%%%%%%%%%%%%%%%%%%%%%%%%%%%%%%%%%%%%%%%%%%%%%%%%%%%%%%%%%%%%%%%%%%%%%%%%%%%%%%%
\section{Analysis} \label{sec:Analysis}
A $C++$ code has been developed alongwith $DISLIN$ code to analyse the equations. $DISLIN$ \cite{dislin} is a freely available graph plotting routine that plots during runtime and can be stored.
\begin{figure}[!h]
    \centering
    \includegraphics[width=0.95\linewidth, height=0.32\textheight]{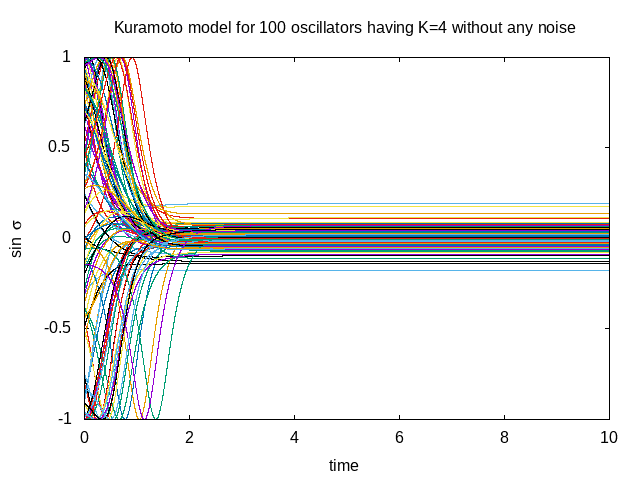}
    \caption{Kuramoto model in arbitrary unit for $100$ oscillators with $K$=$4$ showing synchronization after a certain settling time within a band of frequency range.}
    \label{fig:kuramoto-arbitrary}
\end{figure}
In this section, let us first investigate basic Kuramoto model as discussed in \eqref{eq:basicKura1} including the effect of coupling strength ($K$). If $K$ is properly tuned, one may expect synchronization as shown in Fig.\ref{fig:kuramoto-arbitrary}.

Here we consider that the oscillators are oscillating possessing a frequency distribution $g(\omega)$. One may control width of the distribution while keeping the zero mean. % at certain desired frequency $\omega$.
\begin{figure}[!h]
    \centering
    \includegraphics[width=1.0\linewidth, height=0.32\textheight,
    trim=4 4 4 4, clip]{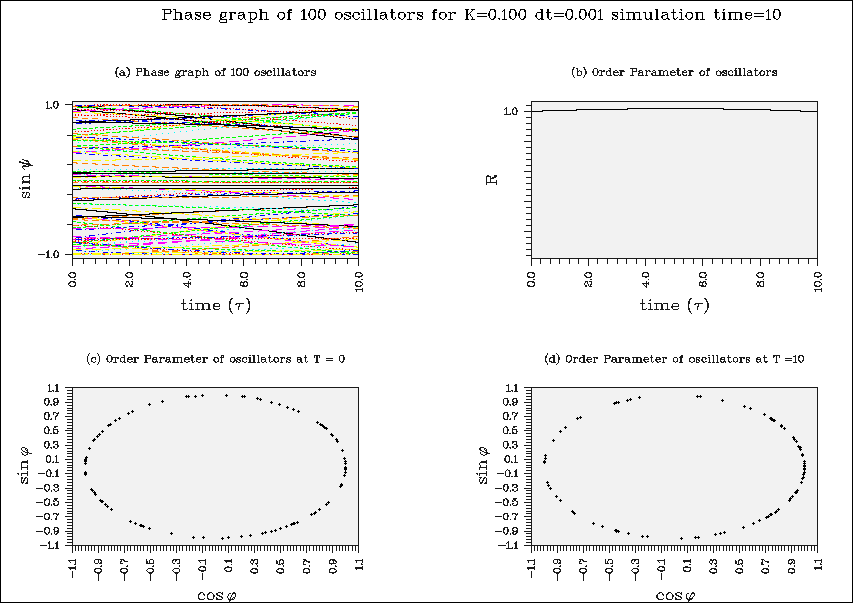}
    \caption{$100$ oscillators with $K$=$0.1$ having Logistic distribution of width $0.001$.}
    \label{fig:n100_0K1Logi001}
\end{figure}

We consider $Logistic$ and $Lorentzian$ fuctions having width $\beta$. Oscillators tend get to be synchronized if $K$ is equal to or more than some threshold value $K_c$ as discussed in \eqref{eq:kc}.
\begin{figure}[!h]
    \centering
    \includegraphics[width=1.0\linewidth, height=0.32\textheight,
    trim=4 4 4 4, clip]{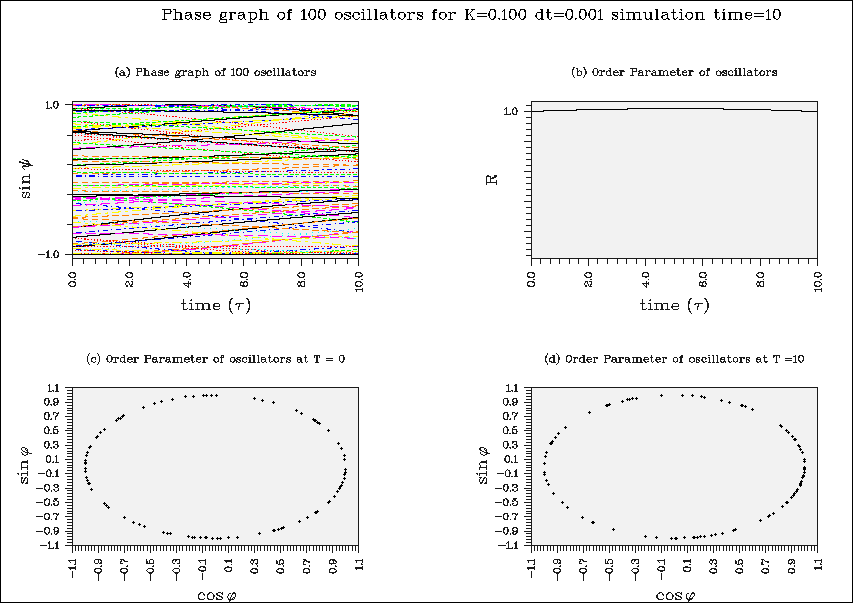}
    \caption{$100$ oscillators with $K$=$0.1$ having Lorentzian distribution of width $0.001$.} \label{fig:n100_0K1Lore001}
\end{figure}
\begin{equation}
g(\omega) = \frac{\exp \left(-\omega / \beta\right)}{\beta \left[1+\exp \left(-\omega / \beta\right)\right]^2} \label{eq:logiEq}
\end{equation}
The Logistic function is described as \eqref{eq:logiEq} which shows $g(0)$=$1/(4 \beta)$ where, $\beta$ is the width. Likewise, one may define the Lorentzian function as \eqref{eq:loreEq}
\begin{equation}
g(\omega) = \frac{b}{\left(\omega^2 + b^2\right)}, \label{eq:loreEq}
\end{equation}
so that one get $g(0)$=$2/(\pi b)$. This $g(0)$ estimates threshold value of the coupling strength as $K_c$=$2/\pi g(0)$.
\begin{figure}[!h]
    \centering
    \includegraphics[width=1.0\linewidth, height=0.32\textheight,
    trim=4 4 4 4, clip]{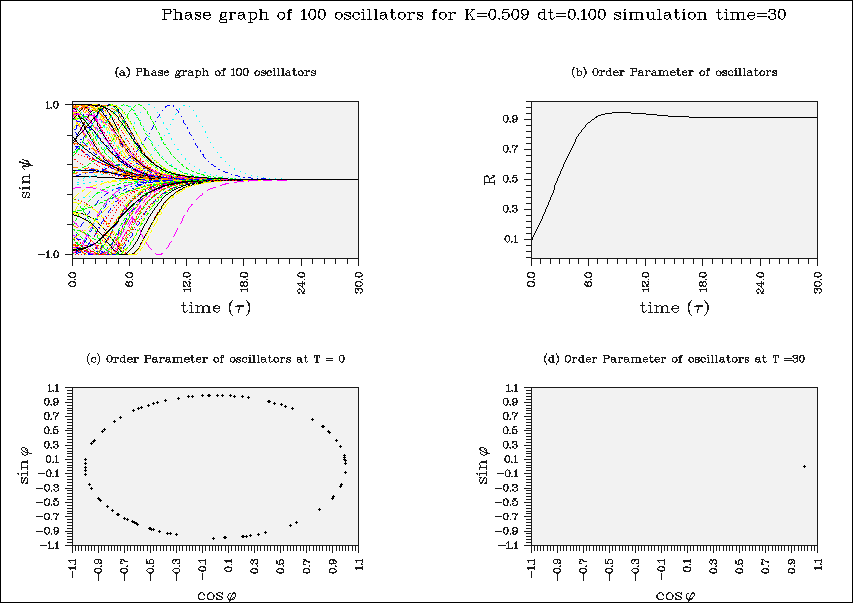}
    \caption{$100$ oscillators oscillating with $k$=$K_c$=$0.509$ with Logistic function of width $0.2$.}
    \label{fig:100_0k509Logi0.20}
\end{figure}
\begin{figure}[!h]
    \centering
    \includegraphics[width=1.0\linewidth, height=0.32\textheight,
    trim=4 4 4 4, clip]{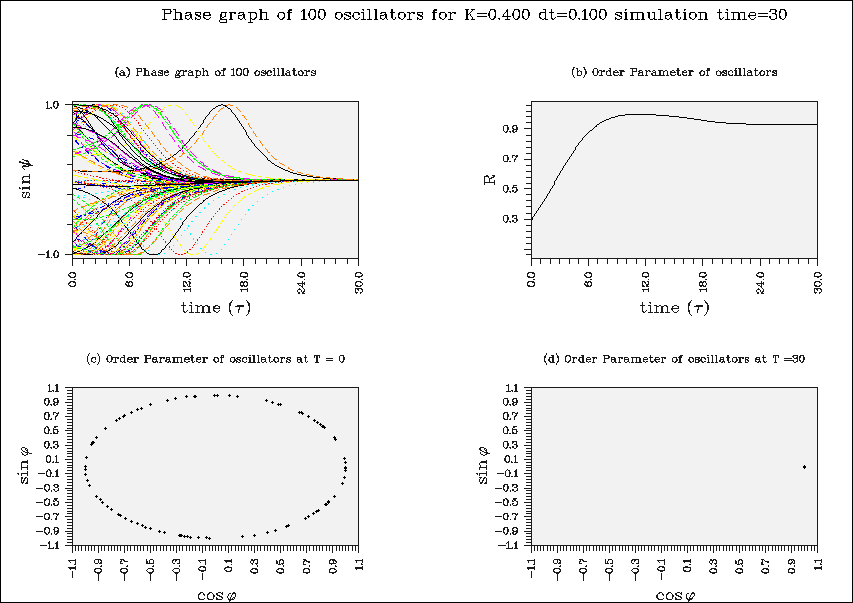}
    \caption{$100$ oscillators oscillating with $k$=$K_c$=$0.4$ with Lorentzian function of width $0.2$.}
    \label{fig:100_0k4Lore0.20}
\end{figure}
One may compare Fig.\ref{fig:n100_0K1Logi001} with Fig.\ref{fig:100_0k509Logi0.20} where the latter is operating with threshold coupling. The synchronization for the latter shows phase space of order parameter as a dot denoting synchronization. Figs.\ref{fig:n100_0K1Lore001} and \ref{fig:100_0k4Lore0.20} also show similar observation of synchronization.

This theoretical study clearly heps us to understand the significance of coupling strength and the treatment of frequency range of oscillators to start with.

\begin{figure}[!h]
    \centering
    \includegraphics[width=1.0\linewidth, height=0.32\textheight,
    trim=4 4 4 4, clip] {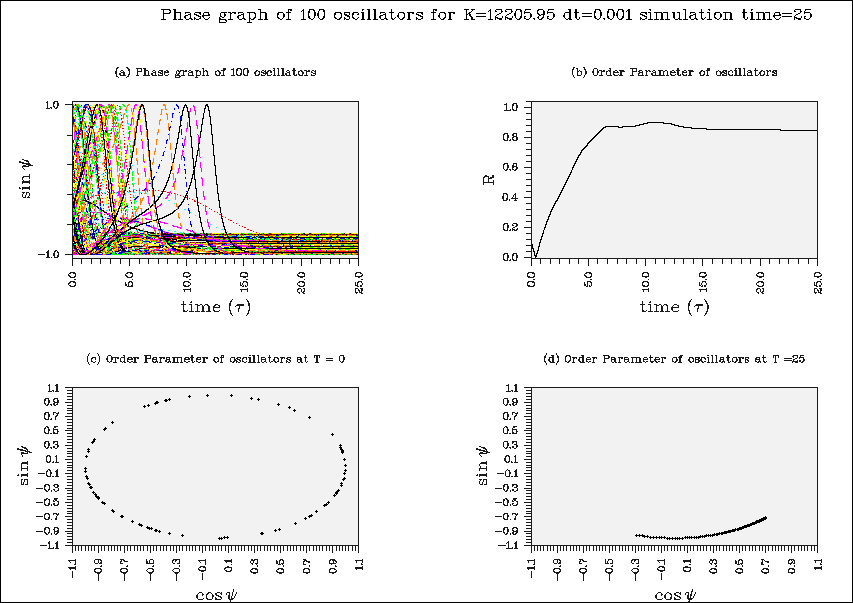}
    \caption{$100$ identical Josephson junctions operating at mean frequency of $5$ $GHz$ having mean $I_c$ = $10$ $\mu A$, internal resistance $\rho$ = $4.2$ $k \Omega$ connected in series array to external load with parameters $L$=$1$ $nH$, $C$ = $1$ $\mu F$ and $R$ = $2$ $\Omega$ treated with bias current $I_b$ = $12$ $\mu A$ synchronizes within a narrow band of distribution. The final phase space is not a dot!}
    \label{fig:100JJI_1}
\end{figure}

Next one may apply this understanding in the case of Josephson junction. The situation is very much different hereas the definition of $K$ is complex for both non-identical and identical junction arrays as evident from either \eqref{eq:psiKuramotoGeneral} or \eqref{eq:psiFinalIdentical2} respectively. Let us consider mean frequency
\begin{figure}[!h]
    \centering
    \includegraphics[width=1.0\linewidth, height=0.32\textheight,
    trim=4 4 4 4, clip] {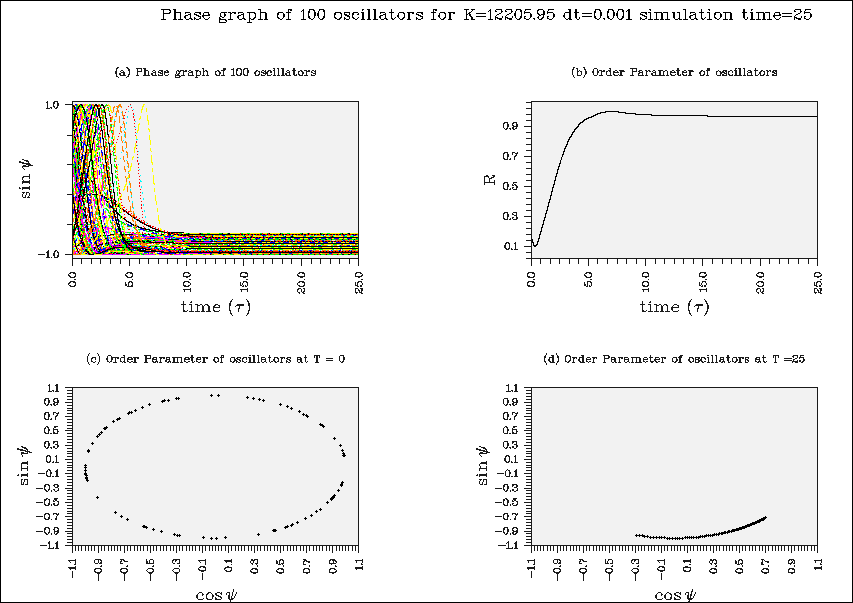}
    \caption{$100$ non-identical Josephson junctions operating with mean frequency of $5$ $GHz$ having mean $I_c$ = $10$ $\mu A$, mean internal resistance $\rho_j$ = $4.2$ $k \Omega$ connected in series array to external load with parameters $L$=$1$ $nH$, $C$ = $1$ $\mu F$ and $R$ = $2$ $\Omega$ treated with bias current $I_b$ = $12$ $\mu A$ synchronizes within a narrow band of distribution.  The final phase space is not a dot!}
    \label{fig:100JJNI_1}
\end{figure}
may be around $5$ $GHz$. Figs.\ref{fig:100JJNI_1} and \ref{fig:100JJI_1} show simulated results of systems

of $100$ Josephson junctions in non-identical and identical configurations respectively operated for $\tilde{\tau}$ = $25$. The interesting part is that synchronization is not pulling the oscillators to a certain unique frequency. Rather, oscillators tend to cool down to a narrow band of frequencies resulting in an arc in phase space diagram which resembles as if oscillators have a certain `viscosity' in the combined system. For the non identical case, the spread of $I_c$ is considered very small like $0.1\%$ while variation in $\rho_j$ is about $0.05$ $\%$ as fabrication is much better and junctions fabricated in the same substrate will not vary too much. Another point to note is that the oscillators in the non-identical case tend to syncronize faster and better than the other case, possibly due to the noisy environment.
\begin{figure}[!h]
    \centering
    \includegraphics[width=1.0\linewidth, height=0.32\textheight,
    trim=4 4 4 4, clip] {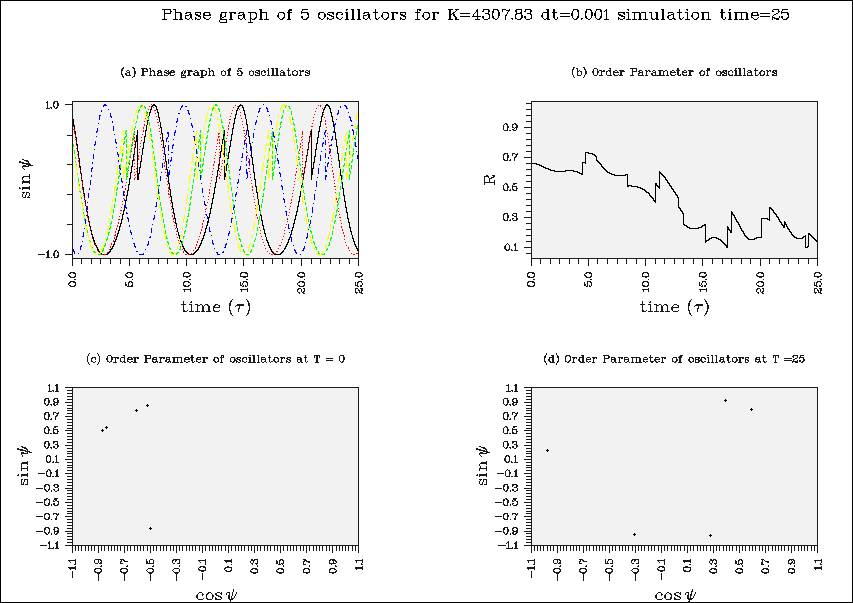}
    \caption{$5$ identical oscillators having $I_c$ = $10$ $\mu A$ and $\rho$ = $4.2$ $k \Omega$ operating with $5$ $GHz$ frequency. } \label{fig:5JJIAsync}
\end{figure}

It has already been discussed that Kuramoto model stands on the assumption that a large number of oscillators have been considered. In our experimental regime, one may need to use smaller number of oscillators say $5$ or $10$ oscillators as shown in Fig.\ref{fig:5JJIAsync} as asynchronized. The order parameter $R$ is also shown to be oscillating at a lower value. The observation was made for $\tilde{\tau}$ = $25$. The circuit parameters were kept same as those for $100$ oscillators. Evidently oscillators were not syhronized. The case for the $5$ non-identical oscillators is same as \ref{fig:5JJIAsync}.
\begin{figure}[!h]
    \centering
    \includegraphics[width=1.0\linewidth, height=0.32\textheight,
    trim=4 4 4 4, clip] {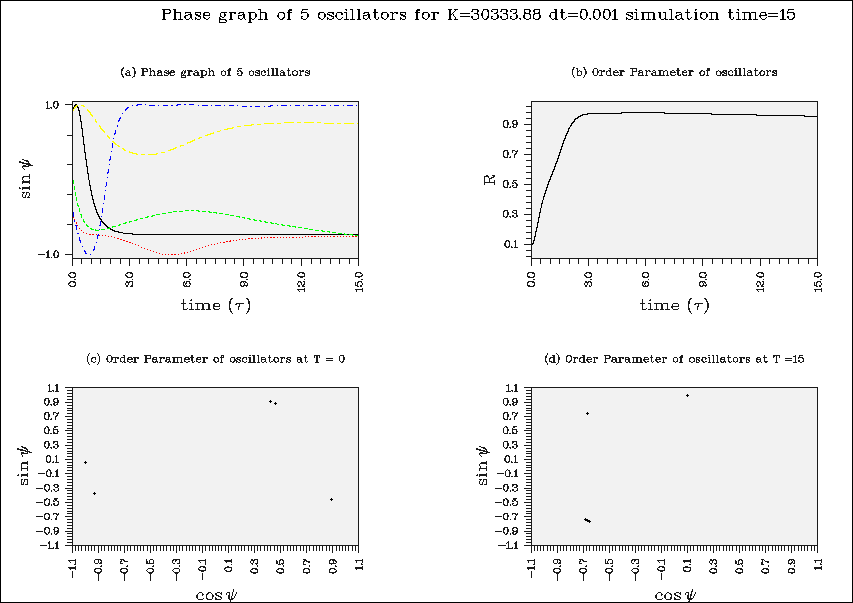}
    \caption{$5$ non identical Josephson junctions are partially syncronized changing $I_b$ to $10.8785$ $\mu A$.}
    \label{fig:5JJNI1}
\end{figure}
Now, to tune the circuit, let us select $I_c$ as $10$ $\mu A$ and $\rho_j$ = $4.2$ $k \Omega$ as before as we wish to experiment with the same junctions while we change $I_b$ - the bias current. In the Fig.\ref{fig:5JJNI1}, the synchronization is observed where one oscillator is out of sync while the rest 4 oscillators come closer to lie in a band very fast $\tilde{\tau}$ $\approx$ $1$.
\begin{figure}[!h]
    \centering
    \includegraphics[width=1.0\linewidth, height=0.32\textheight,
    trim=4 4 4 4, clip] {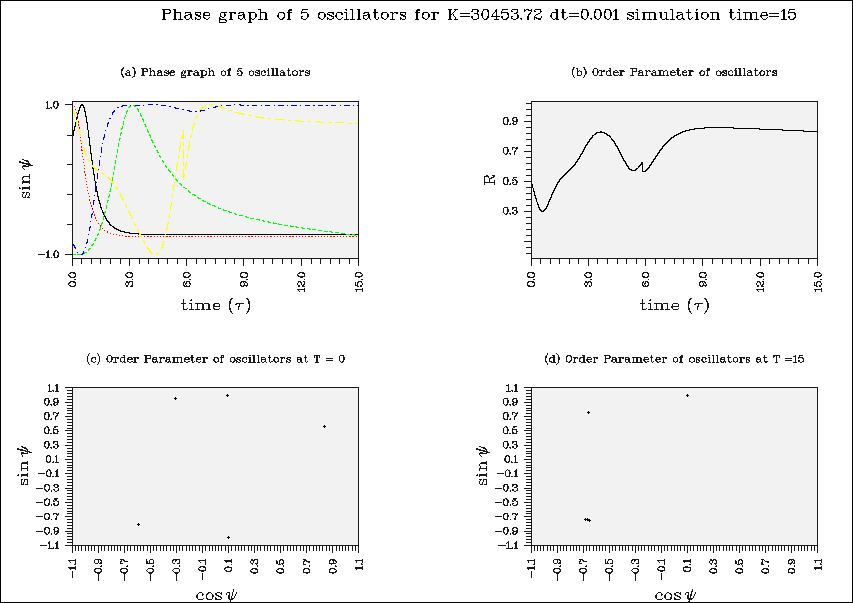}
    \caption{$5$ identical Josephson junctions are partially syncronized changing $I_b$ to $10.877$ $\mu A$.}
    \label{fig:5JJI1}
\end{figure}

\section{Conclusion}
The exercises demonstrated in Figs.\ref{fig:5JJNI1} and \ref{fig:5JJI1} show the possibility of synchronization for few oscillators following Kuramoto model. However, order parameter show in-course instability which later settles down.

This study helps to understand applicability of junctions in series array and steps to control the level of synchronization. The process is easier and synchronization is performed well for larger number of junctions while partial synchronization is also possible following the Kuramoto model. However, this study does not state any conclusive equation for threshold coupling for Josephson junction as it discussed in case of general oscillators. This aspect will be discussed in future.

%%%%%%%%%%%%%%%%%%%%%%%%%%%%%%%%%%%%%%%%%%%%%%%%%
\appendices
\section{} \label{sec:appendix1}
From \eqref{eq:naturalAnglexfrm},
\begin{eqnarray}
& & \tan \left( \frac{\phi_j}{2} + \frac{\pi}{4} \right) = \sqrt{\frac{\alpha_j+1}{\alpha_j-1} } \tan \frac{\psi_j}{2},   \nonumber \\
& \text{or,}& \left\{ \tan \left( \frac{\phi_j}{2} + \frac{\pi}{4} \right) + 1\right\} \left\{ \tan \left( \frac{\phi_j}{2} + \frac{\pi}{4} \right) - 1\right\}  \nonumber \\
& & = \left\{ \sqrt{\frac{\alpha_j+1}{\alpha_j-1} }   \tan \frac{\psi_j}{2}+1\right\} \left\{\sqrt{\frac{\alpha_j+1}{\alpha_j-1} }   \tan \frac{\psi_j}{2}-1\right\}, \nonumber  \\
&\text{or,}& \left(\frac{1+\tan \frac{\phi_j}{2}}{1- \tan{\frac{\phi_j}{2}}} + 1 \right) \left(\frac{1+\tan \frac{\phi_j}{2}}{1- \tan{\frac{\phi_j}{2}}} - 1 \right)  \nonumber \\
& & =  \frac{\alpha_j+1}{\alpha_j-1} \tan^2 \frac{\psi_j}{2}  -1, \nonumber \\
& \text{or,}& \left( \frac{\cos \frac{\phi_j}{2}+\sin \frac{\phi_j}{2}}{\cos \frac{\phi_j}{2}-\sin \frac{\phi_j}{2}} + 1 \right) \left( \frac{\cos \frac{\phi_j}{2}+\sin \frac{\phi_j}{2}}{\cos \frac{\phi_j}{2}-\sin \frac{\phi_j}{2}} - 1 \right) \nonumber \\
& & =  \frac{\alpha_j+1}{\alpha_j-1} \tan^2 \frac{\psi_j}{2}  -1, \nonumber \\
&\text{or,}& \frac{2 \cos \frac{\phi_j}{2} \times 2 \sin \frac{\phi_j}{2}}{\left(\cos \frac{\phi_j}{2}-\sin \frac{\phi_j}{2}\right)^2} = \frac{2 \sin \phi_j}{\left(\cos \frac{\phi_j}{2}-\sin \frac{\phi_j}{2}\right)^2}  \nonumber \\
& & = \frac{\alpha_j+1}{\alpha_j-1} \tan^2 \frac{\psi_j}{2}  -1, \nonumber \\
& \text{or, }& \frac{2 \sin \phi_j}{1-\sin \phi_j} = \frac{\alpha_j \tan^2 \frac{\psi_j}{2} + \tan^2 \frac{\psi_j}{2} - \alpha_j +1}{\alpha_j-1},  \nonumber \\
%\end{eqnarray}
%\begin{eqnarray}
& \text{or, }& \frac{1-\sin \phi_j}{2 \sin \phi_j} = \frac{1-\alpha_j \cos \psi_j}{\left(\alpha_j-1\right) \cos^2 \frac{\psi_j}{2}}, \nonumber \\
& \text{or, }& \frac{1}{2 \sin \phi_j}  =  \frac{1}{2} +  \frac{1-\alpha_j \cos \psi_j}{\left(\alpha_j-1\right) \cos^2 \frac{\psi_j}{2}}, \nonumber \\
&\text{or, }& \sin \phi_j = \frac{1- \alpha_j \cos \psi_j}{\alpha_j - \cos \psi_j}, \nonumber \\
& \text{or, }& \sin \phi_j \equiv \sin \phi(\psi_j) = \alpha_j - \frac{\left(\alpha_j^2-1 \right)}{\alpha_j - \cos \psi_j}. \nonumber
\end{eqnarray}

\section{} \label{sec:appendix2}
\begin{eqnarray}
& & \left \langle \frac{d \psi_j}{d\tilde{\tau}}\right \rangle =  1 - \frac{\epsilon_j \delta_j \sqrt{\alpha_j^2-1}}{2 \pi N} \int_0^{2 \pi} \left(\frac{\alpha_j - \cos \psi_j}{\alpha_j^2 - 1} \right.\nonumber \\
& &\times \left.\sum_{k=1}^N \sum_{n=0}^\infty n  I_k \rho_k B_{kn} \sin \left\{n \left(\tilde{\tau} + c_k\right) + \beta_{kn}\right\} \right) d\tilde{\tau}. \nonumber \\
&\text{or, }& \left \langle \frac{d \psi_j}{d\tilde{\tau}}\right \rangle = 1 - \frac{\epsilon_j \delta_j }{2 \pi N \sqrt{\alpha_j^2-1}} \int_0^{2 \pi} \left(\alpha_j - \cos \left(\tilde{\tau} + c_j\right) \right)\nonumber \\
& &\times \sum_{k=1}^N \sum_{n=0}^\infty n  I_k \rho_k B_{kn} \sin \left\{n \left(\tilde{\tau} + c_k\right) + \beta_{kn}\right\}d\tilde{\tau}. \nonumber \\
&\text{or, }& \left \langle \frac{d \psi_j}{d\tilde{\tau}}\right \rangle = 1 \nonumber \\
& & + \frac{\epsilon_j \delta_j}{N \sqrt{\alpha_j^2-1}\sqrt{\gamma_j^2 \left(\alpha_j^2-1\right)^2 + \left(\omega_{0j}^2-\left(\alpha_j^2-1\right)^2\right)^2}}\nonumber \\
& &\times \sum_{k=1}^N I_k \rho_k \left(1-\alpha_k^2+\alpha_k\sqrt{\alpha_k^2-1}\right) \sin \left(cj-c_k-\zeta_j \right). \nonumber \\
&\text{or, }& \left \langle \frac{d \psi_j}{d\tilde{\tau}}\right \rangle \nonumber \\
& & = 1 +\frac{K_j}{N} \sum_{k=1}^N I_k \rho_k \left(1-\alpha_k^2+\alpha_k\sqrt{\alpha_k^2-1}\right) \nonumber \\
& & \times \sin \left(c_j-c_k-\zeta_j \right), \nonumber \\
&\text{or, }& \left \langle \frac{d \psi_j}{d\tilde{\tau}}\right \rangle = 1 + \frac{K_j}{N} \sum_{k=1}^N A_k \sin \left(c_j - c_k - \zeta_j \right). \nonumber
\end{eqnarray}

%%%%%%%%%%%%%%%%%%%%%%%%%%%%%%%%%%%%%%%%%%%%%%%%%%%%%%%%%%%%%%%%%%%%%%%%%%%%%%%%%
\section*{Acknowledgment}
The author would like to Sudhir R Jain, for his ideas, inspiration and continuous support to conceptualize, understand and formulate the problem. The author also expresses gratitude to Susmita Bhattacharyya and Tilottoma Bhattacharyya for their guidance.
%%%%%%%%%%%%%%%%%%%%%%%%%%%%%%%%%%%%%%%%%%%%%%%%%%%%%%%%%%%%%%%%%%%%%%%%%%%%%%%%%%%%%

%%%%%%%%%%%%%%%%%%%%%%%%%%%%%%%%%%%%%%%%%%%%%%%%%%%%%%%%%%%%%%%%%%%%%%%%%%%%%%%%%
\bibliography{Kuramoto_Paper}

% Generated by IEEEtran.bst, version: 1.14 (2015/08/26)
\begin{thebibliography}{10}
\providecommand{\url}[1]{#1}
\csname url@samestyle\endcsname
\providecommand{\newblock}{\relax}
\providecommand{\bibinfo}[2]{#2}
\providecommand{\BIBentrySTDinterwordspacing}{\spaceskip=0pt\relax}
\providecommand{\BIBentryALTinterwordstretchfactor}{4}
\providecommand{\BIBentryALTinterwordspacing}{\spaceskip=\fontdimen2\font plus
\BIBentryALTinterwordstretchfactor\fontdimen3\font minus
  \fontdimen4\font\relax}
\providecommand{\BIBforeignlanguage}[2]{{%
\expandafter\ifx\csname l@#1\endcsname\relax
\typeout{** WARNING: IEEEtran.bst: No hyphenation pattern has been}%
\typeout{** loaded for the language `#1'. Using the pattern for}%
\typeout{** the default language instead.}%
\else
\language=\csname l@#1\endcsname
\fi
#2}}
\providecommand{\BIBdecl}{\relax}
\BIBdecl

\bibitem{benz2004}
S.~P. Benz and C.~A. Hamilton, ``Application of josephson effect to voltage
  metrology,'' \emph{Proceedings of the IEEE}, vol.~92, no.~10, pp. 1617--1629,
  2004.

\bibitem{josephson1961}
B.~D. Josephson, ``Possible new effects in supercondictive tunneling,''
  \emph{Physics Letters}, vol.~1, no.~7, pp. 251--253, 1962.

\bibitem{josephson1964}
------, ``Coupled superconductors,'' \emph{Reiew of Modern Physics}, vol.~36,
  no.~1, pp. 216--220, 1964.

\bibitem{josephson1973}
------, ``The discovery of tunneling supercurrents,'' \emph{Nobel Lectures},
  1973.

\bibitem{deaver1961}
B.~S.~D. Jr. and W.~M. Fairbank, ``Experimental evidence for quantized flux in
  superconducting cylinders,'' \emph{Physical Review Letters}, vol.~7, no.~2,
  pp. 43--46, 1961.

\bibitem{endo1983}
T.~Endo, masao Koyanagi, and A.~Nakamura, ``High accuracy josephson
  potentiometer,'' \emph{IEEE Transactions on Instrumentation and Measurement},
  vol. IM-32, no.~1, pp. 267--271, 1983.

\bibitem{levinsen1977}
M.~T. Levinsen, R.~Y. Chiao, M.~J. Feldman, and B.~A. Tucker, ``Applied physics
  letters,'' \emph{Applied Physics Letters}, vol.~31, p. 776, 1977.

\bibitem{benz1991}
S.~P. Benz and C.~J. Burroughs, ``Coherent emission from two dimensional
  josephson junction arrays,'' \emph{Applied Physics Letters}, vol. 58(19), pp.
  2162--2164, 1991.

\bibitem{wan1989}
K.~Wan, A.~K. Jain, and J.~E. Lukens, ``Submillimeter wave generation using
  josephson junction arrays,'' \emph{Applied Physics Letters}, vol.~54, pp.
  1805--1807, 1989.

\bibitem{swift1992}
J.~W. Swift, S.~H. Strogatz, and K.~Wisenfield, ``Averaging of globally coupled
  oscillators,'' \emph{Physica D}, vol.~55, pp. 239--250, 1992.

\bibitem{swift1995}
K.~Wisenfeld and J.~W. Swift, ``Averaged equations for josephson junction
  series arrays,'' \emph{Physical Review E}, vol.~51, pp. 1020--1025, 1995.

\bibitem{kuramoto1986}
H.~Sakaguchi and Y.~Kuramoto, ``Kuramoto order parameters and phase
  concentration for the kuramoto-sakaguchi equation with frustration,''
  \emph{Progress of Theoretical Physics}, vol. 76(3), pp. 576--581, 1986.

\bibitem{guckenheimer2002}
J.~Guckenheimer and P.~Holmes, \emph{Nonlinear Oscillations, Dynamical Systems,
  and Bifurcations of Vector fields}.\hskip 1em plus 0.5em minus 0.4em\relax
  Springer Verlag New York, 2002.

\bibitem{ha2021}
S.-Y. Ha, J.~Morales, and Y.~Zhang, ``A soluble active rotator model showing
  phase transitions via mutual entrainment,'' \emph{Communications on pure and
  applied analysis}, vol. 20(7 \& 8), pp. 2579--2612, 2021.

\bibitem{dorfler2011}
D.~Florian and F.~Bullo, ``On the critical coupling for kuramoto oscillators,''
  \emph{SIAM Journal of Apllied Dynamical Systems}, vol.~10, 2011.

\bibitem{dislin}
\BIBentryALTinterwordspacing
H.~Michels, ``Dislin software.'' [Online]. Available:
  \url{http://www.dislin.de/}
\BIBentrySTDinterwordspacing

\end{thebibliography}
\bibliographystyle{IEEEtran}

\end{document}